\documentclass[10pt]{article}
\usepackage{authblk} 
\usepackage{color}
\usepackage{refcount}
\usepackage{graphicx}
\usepackage{slashed}
\usepackage{braket,amsmath,amssymb}
\usepackage[export]{adjustbox}
\usepackage{subcaption}
\usepackage{graphics,psfrag,empheq,float}
\usepackage[colorlinks=true,linkcolor=black]{hyperref}
\usepackage{cite}
\usepackage{wasysym}
\usepackage[normalem]{ulem} 
\def\XXint#1#2#3{{\setbox0=\hbox{$#1{#2#3}{\int}$}
     \vcenter{\hbox{$#2#3$}}\kern-.5\wd0}}

\newcommand{\vp}{{\boldsymbol{p}}}
\newcommand{\vq}{{\boldsymbol{q}}}

\newcommand{\Ima}{{\rm Im}}

\newcommand{\sgn}{{\rm sgn}}

\newcommand{\bg}{\begin{align}}
\newcommand{\eeg}{\end{align}}
\newcommand{\be}{\begin{equation}}
\newcommand{\ee}{\end{equation}}
\newcommand{\ba}{\begin{eqnarray}}
\newcommand{\ea}{\end{eqnarray}}

\newcommand{\nn}{\nonumber}

\newcommand{\vep}{\varepsilon}

\newcommand{\ra}{\rangle}

\newcommand{\ep}{\epsilon}

\newcommand{\lam}{\lambda}
\newcommand{\lm}{\lambda}

\newcommand{\tiam}{{\text{IAM}}}
\newcommand{\tnd}{{\text{ND}}}
\newcommand{\tndl}{\text{ND0}}
\newcommand{\tndn}{\text{ND1}}

\newcommand{\jo}[1]{\textcolor{black}{{{#1}}}}

\begin{document}

\title{Unitarization of the one-loop graviton-graviton scattering amplitudes and study of the graviball}

\author[a]{J. A. Oller\thanks{oller@um.es}}
\author[b]{Marcela Pel\'aez\thanks{mpelaez@fing.edu.uy}}
\affil[a]{\it Departamento de F\'{\i}sica, Universidad de Murcia, E-30071 Murcia,  Spain}
\affil[b]{{\it 
  Instituto de F\'{\i}sica, Facultad de Ingenier\'{\i}a, Universidad de la Rep\'ublica,} \authorcr {\it J. H. y Reissig 565, 11000 Montevideo, Uruguay } }

\maketitle

\begin{abstract}
  From the graviton-graviton scattering amplitudes calculated perturbatively in quantum gravity to the one-loop order, we develop further a formalism that allows one to calculate infrared-finite partial-wave amplitudes fulfilling  perturbative unitarity. As a result of this process a parameter dubbed $\ln a$ emerges that separate between infrared and typical external momenta. The resulting partial-wave amplitudes are next unitarized by employing  the Inverse Amplitude Method and the algebraic-$N/D$ method. Then, the  graviball resonance, with a similar pole position, is confirmed in the $S$-wave partial-wave amplitude for all unitarization methods, also with respect to the unitarization of only the leading-order amplitude. \jo{Although the spectrum of the theory is independent of the specific value of $\ln a$, the requirement for a well-behaved unitarized effective field theory of gravity identifies the optimal range of $\ln a$ for our next-to-leading-order calculations as $0.5 \lesssim \ln a \lesssim 1.7$.}  Briefly, we discuss the $D$-wave scattering that is weaker than the $S-$wave scattering, repulsive and non-resonant  for $\ln a\approx 1$.
\end{abstract}


\newpage
\section{Introduction}
\def\theequation{\arabic{section}.\arabic{equation}}
\setcounter{equation}{0}

In this work, we study graviton-graviton scattering by unitarizing the one-loop perturbative four-graviton scattering amplitudes calculated in perturbative gravity in Ref.~\cite{Dunbar:1994bn}.  These scattering amplitudes are ultraviolet finite and lack of any counterterm \cite{vanNieuwenhuizen:1976vb,AccettulliHuber:2019jqo} but they are infrared divergent \cite{Dunbar:1994bn,Dunbar:1995ed,Donoghue:1999qh}. The one-loop scattering amplitudes are suppressed by an extra power of $Gs$ compared to the tree-level amplitudes for low-energy gravity, $|s|\ll G^{-1}$, being $s$ the total center of mass energy squared and $G$ the Newton constant. This energy region is typically the realm of the low-energy effective-field theory (EFT) of gravity \cite{Donoghue:2017pgk,Weinberg:1978kz,Donoghue:1993eb,Donoghue:1994dn,Bjerrum-Bohr:2002gqz,Burgess:2003jk}. The unitarization process has to face the intrinsic difficulties associated to the infrared-divergent character of the scattering amplitudes. Here, we take into account that 
it is possible to tackle the infrared divergences so as to end with finite $S$-matrix amplitudes. In recent years, the developing of a well-defined $S$-matrix theory for gravity as well as for  QED with massless charged particles and in Yang-Mills theory has experienced  striking advances in 
different directions \cite{Ware:2013zja,Choi:2017bna,Choi:2017ylo,Prabhu:2022zcr,Prabhu:2024zwl,Prabhu:2024lmg,Campiglia:2016hvg,Campiglia:2014yka,Ashtekar:2018lor,Hirai:2020kzx,Himwich:2020rro,Hannesdottir:2019umk,Elkhidir:2024izo}.

An infrared finite $S$-matrix for Quantum Electrodynamics (QED) was constructed long time ago by Faddeev and Kulish (FK) in Ref.~\cite{Kulish:1970ut}.  Reference~\cite{Ware:2013zja} has  provided an explicit extension of this method in perturbative gravity, and the properties of the FK states in gravity under the BMS group were extensively studied in Refs.~\cite{Choi:2017bna,Choi:2017ylo}. \jo{In connection with this, Ref.~\cite{Elkhidir:2024izo} shows how BMS supertranslations can also be captured by the on-shell framework of scattering amplitudes.}
In the present work we use the observation of Ref.~\cite{Oller:2024qkw} for Coulomb scattering in QED that one can extract  the same infrared-finite scattering amplitudes as in the FK method by applying the compact formulas provided by Weinberg \cite{Weinberg:1965nx} that resum the infrared divergences from the exchanges of soft virtual photons/gravitons. In this way, one does not need the auxiliary construction of the FK states in order to end with the $S$-matrix elements. Remarkably, this result has been generally proved by Ref.~\cite{Choi:2017bna} to all orders in perturbative gravity.  
Infrared divergences  also make that the standard definition of partial-wave amplitudes (PWAs) cannot be directly implemented.  In particular, the calculation of infrared-finite PWAs in terms of the infrared-finite scattering amplitudes already obtained in momentum space has to be properly done because of diverging integrals in the angular variables.  
In this respect, we follow and expand the ideas in the approach of Refs.~\cite{Blas:2020dyg,Blas:2020och,Oller:2022tmo} for calculating infrared finite PWAs for \jo{infinite}-range forces in PWAs. 

By studying the characteristic  size of one-loop unitarity contributions  and their variation under a change of scale, it was concluded in Ref.~\cite{Blas:2020och,Blas:2020dyg} that a typical scale for the low-energy graviton-graviton  scale is $Q^2=\pi G^{-1}/\ln a$.  \jo{The parameter $\ln a$ emerges during the process of removing  infrared divergences and is related to the cutoff in momentum, $\Lambda^2=s/a^2$, such that  soft(hard) gravitons have linear momenta smaller(larger) than $\Lambda$ in magnitude.}    
We also compare the results between different unitarization methods \cite{Oller:2020guq,Oller:2019opk}: The algebraic $N/D$ method \cite{Oller:1998zr,Oller:1999me}, the first-iterated $N/D$ method \cite{martin.200705.1}, and the Inverse Amplitude Method (IAM) \cite{Lehmann:1972kv,Truong:1988zp,Dobado:1989qm,Delgado:2022uzu}.   Our benchmark results are those obtained by applying the IAM because this method does not introduce any unitarization  scale.  \jo{These unitarization methods are very popular in hadron physics, where these methods, and variants of them,  have been applied with great success for studying many resonances and related processes \cite{Lehmann:1972kv,Truong:1988zp,Oller:2019rej,Oller:2019opk,Oller:2020guq,Oller:2000ma,Mai:2020ltx}.}

Reference~\cite{Blas:2020och} established the existence of a low-lying graviton-graviton resonance in pure gravity, the so-called {\it graviball}, by unitarizing the Born term for graviton-graviton scattering. It corresponds to a resonance pole $s_P$ in the scalar $J=0$ ($J$ is the total angular momentum) PWA, with the value $s_P=(0.22-i\,0.63)(G\ln a)^{-1}$ obtained in Ref.~\cite{Blas:2020och}, so that $|s_P|/Q^2\simeq 0.2 \ll 1$.   Let us notice that the convergence radius of the perturbative series is limited by $|s_P|\ll Q^2$.  
An interesting argument was put forward in Ref.~\cite{Blas:2020dyg} to favor a value of  $\ln a\approx 1$.

This reference highlights the remarkable similarity between the graviball pole and the lightest resonance in QCD, known as the $\sigma$ or $f_0(500)$ \cite{ParticleDataGroup:2022pth}. Notably, the pole position for the $\sigma$ resonance satisfies $|s_\sigma|/(4\pi f_\pi)^2 \simeq 0.2$, where $f_\pi \approx 92$~MeV is the pion decay constant, \jo{and $4\pi f_\pi$ represents the natural scale for the low-energy EFT of QCD, Chiral Perturbation Theory \cite{Weinberg:1966fm,Gasser:1983yg}, which describes pion interactions and other hadronic processes. Since the low-energy limits of quantum gravity and QCD can both be described,  to some extent, by similar EFTs, such analogies are not entirely surprising. The presence of the $\sigma$ pole in the scalar isoscalar $\pi\pi$ PWA can be rigorously established by combining EFT-based amplitude calculations with S-matrix methods \cite{Oller:1997ti,Caprini:2005zr}, or  exclusively by using the latter \cite{Garcia-Martin:2011nna}.}

In our present study, the unitarization of the one-loop scattering amplitudes confirms the $J=0$  graviball pole. 
Nonetheless,  while $|s_P|$ is almost the same, now the real part of $s_P$ increases while its imaginary part decreases,  giving rise to a sharper resonance signal along the real $s$-axis as compared with the LO study \cite{Blas:2020och,Blas:2020dyg}. \jo{We also discuss that the pole content of the infrared-finite $S$-matrix should not depend on a specific value of $\ln a$ for a full calculation to all orders in powers of $G$.} 
  The requirement of having a well-behaved unitarized EFT for graviton-graviton scattering implies interesting  constraints.  Through the application of different unitarization methods and values of $\ln a$,  variations in the results along the real axis, $0<s\ll Q^2$, as well as regarding the graviball pole position, are compared with expected sizes based on power counting. \jo{This leads us to argue that the optimized interval of values of $\ln a$ for our NLO calculations is $0.5\lesssim \ln a\lesssim 1.7$. }

  After this introduction, we explain the main features of the formalism 
  in Sec.~\ref{sec.240517.1}. The unitarization methods considered 
  are the object of Sec.~\ref{sec.240517.2}. We 
 study the $J=0$ graviton-graviton PWA and its spectrum in Sec.~\ref{sec.240517.2}. The main features of the $J = 2$ PWA are also briefly discussed.  Section~\ref{sec.240517.4} ends the manuscript  with some concluding remarks and outlook.

\section{Formalism: Infrared-finite scattering amplitudes and partial-wave amplitudes}
\def\theequation{\arabic{section}.\arabic{equation}}
\setcounter{equation}{0}   
\label{sec.240517.1}

For the generic scattering process $\alpha\to \beta$,  the relation between the full $S$-matrix $S_{\beta\alpha}$, and the one without infrared divergences $S_{\beta\alpha}^0$,  was  established by Weinberg in \cite{Weinberg:1965nx} for gravity and QED. By employing dimensional regularization (DR) we would write
\begin{align}
  \label{240423.1a}
  S_{\beta\alpha}&=S_{\beta\alpha}^0\exp\left(\frac{1}{2}\int^\Lambda d^D q B(q)\right)~,
  \end{align}
where $D=4-2\epsilon$, \jo{and  the integration variables correspond to the components of $q$, which is the momentum of and exchanged soft graviton. Here,  $\Lambda$ is  a cutoff that separates between soft, $|\vq|<\Lambda$, and hard, $|\vq|>\Lambda$, gravitons \cite{Weinberg:1965nx}.} The exponent in Eq.~\eqref{240423.1a} is denoted by $\Phi_{\beta \alpha}$ and it is given by the expression \cite{Weinberg:1965nx} (in our convention the Minkowski metric is $\text{diag}(-1,+,+,+)$), 
\begin{align}
\label{240517.1}
\Phi_{\beta\alpha}&\equiv \frac{1}{2}\int d^D q B(q)=-i 4\pi G \sum_{n,m}\eta_n\eta_m \int^\Lambda \frac{d^Dq}{(2\pi)^D}\frac{(p_n\cdot p_m)^2-\frac{1}{2}m_n^2m_m^2}{(q^2+i\vep)(p_n\cdot q+i\eta_n \vep)(p_m\cdot q-i\eta_m \vep)}~,
\end{align}
\jo{with $\vep\to 0^+$, and the indices $m$ and $n$ run both over all the particles involved in the initial and final states. In this equation the mass and momentum of the $n_{\text{th}}$ particle are designed by $m_n$ and $p_n$ ($p_n^2=m_n^2$), respectively. The sign function}  $\eta_n=+1(-1)$ for out(in)going particles. 

We have performed \jo{the} calculation \jo{of $\Phi_{\beta\alpha}$} within DR, since  graviton-graviton scattering amplitudes from  Ref.~\cite{Dunbar:1994bn} are calculated with this regularization scheme. \jo{Dimensional regularization regularizes the result for the $q\to 0$ part of the integral in Eq.~\eqref{240517.1}.}  More details of its calculation will be given elsewhere \cite{Oller:2024nss}.   The result \jo{in the massless limit ($m_n\to 0$)} is
\begin{align}
\label{240424.7}
\Phi_{\beta\alpha}&=\frac{1}{2\pi}\left(\frac{1}{\hat{\ep}}-\ln\Lambda^2\right)\sum_{n,m}\eta_n\eta_m p_n\cdot p_m\ln(-\eta_n\eta_m 2 p_n\cdot p_m)~,
\end{align}
with $\frac{1}{\hat\epsilon}=\frac{1}{\ep}+\ln 4\pi-\gamma_E$. It also follows from Eq.~\eqref{240423.1a} that the $S$ matrix free of infrared divergences $S_{\beta\alpha}^0$ corresponds to $ S^0_{\beta\alpha}=S_{\beta\alpha}\exp\left(-\Phi_{\beta\alpha} \right)$. \jo{Since $S_{\beta\alpha}$ does not depend on $\Lambda$, Eqs.~\eqref{240423.1a} and \eqref{240424.7} imply that the two infrared finite $S$ matrices calculated with the cutoffs $\Lambda$ and $\Lambda'$ are related by the factor}
\begin{align}
  \label{241222.1}
  \frac{S^0_{\beta\alpha}(\Lambda)}{S^0_{\beta\alpha}(\Lambda')}=\exp\left(\frac{1}{\pi}\ln\frac{\Lambda}{\Lambda'}\sum_{n,m}\eta_n\eta_m p_n\cdot p_m\ln(-\eta_n\eta_m 2 p_n\cdot p_m)
  \right)\,.
\end{align}
The relation between the $S$ and $T$ matrices is $S^0_{\beta\alpha}=\delta(\beta-\alpha)+i T_{\beta \alpha}$. The graviton-graviton scattering amplitude $T_{\beta\alpha}$ in the loop expansion of gravity as a low-energy EFT is written as
\begin{align}
\label{240423.2}
T_{\beta\alpha}&=T_{\beta\alpha}^{(0)}+T_{\beta\alpha}^{(1)}+\ldots
\end{align}
\jo{where the} superscript between brackets indicates the number of loops involved in the corresponding scattering amplitude, and the ellipsis represents higher orders. We consider here contributions up to the one-loop order or ${\cal O}(G^2)$.  Then, the $T$ matrix up to ${\cal O}(G^2)$ gets dressed as
\begin{align}
  \label{240423.3}
 T^{(0)}_{\beta\alpha}+T^{(1)}_{\beta\alpha}-T^{(0)}_{\beta\alpha}\,\Phi_{\beta\alpha}~. 
\end{align}
As indicated in the introduction, when applied to QED this procedure is equivalent to the FK method \cite{Kulish:1970ut,Oller:2024qkw}, and  a proof to all orders in perturbative gravity is given in Ref.~\cite{Choi:2017bna}.  
Moreover, there is  another  contribution that stems from the identity operator in the $S$ matrix times the expansion of the phase factor,
\begin{align}
  \label{240423.5}
\mathbb{I}\,\exp\left(-i\,\Ima \Phi_{\beta\alpha}\right) .
\end{align}
The real part of $\phi_{\beta\alpha}$ does not enter in the dressing the identity operator because the normalization of the states is fixed. \jo{This is explicitly demonstrated  for two-body scattering below Eq.~\eqref{240425.11}.}  
Equation~\eqref{240423.5}  is essential  for the calculation of infrared finite PWAs \cite{Blas:2020och,Blas:2020dyg},  because the angular integration over the whole scattering solid angle picks up the identity operator contribution.

\subsection{Explicit formulas for the infrared-finite scattering amplitudes}

We specifically consider the graviton-graviton scattering process 
$|\vp_1\lam_1\ra|\vp_2\lam_2\ra\to|\vp_3\lam'_1\ra|\vp_4\lam'_2\ra$, 
where $|\vp_i\lam_i\ra$ denotes a graviton state with linear momentum $\vp_i$ and helicity $\lam_i$. 
We denote the scattering amplitudes directly calculated from perturbative gravity by $T_{\lambda'_1\lambda_2';\lambda_1\lambda_2}(s,t,u)$, where $s$, $t$ and $u$ are the usual Mandelstam variables: $s=(p_1+p_2)^2$, $t=(p_1-p_3)^2$ and $u=(p_1-p_4)^2$. In turn, the infrared-finite graviton-graviton scattering amplitudes are denoted as $A_{\lambda_1'\lambda_2';\lambda_1\lambda_2}(s,t,u)$.  Since the helicity of a graviton is $\pm 2$,  we will only indicate its sign for simplicity. 
We can express the different graviton-graviton scattering amplitudes 
 just in terms of three of them, being the others related by crossing and $P$ and $T$ invariance. At the tree level we have \cite{Dunbar:1994bn}, 
\begin{align}
\label{240518.1}
T^{(0)}_{++;++}(s,t,u)&=\frac{8\pi Gs^4}{stu}~,\\
 \label{240425.5}
T^{(0)}_{+-;++}(s,t,u)&=0~,\\
T^{(0)}_{--;++}(s,t,u)&=0~.\nn
\end{align}
At the one-loop level the expressions are \cite{Dunbar:1994bn}, 
\begin{align}
\label{240518.2}
T^{(1)}_{++;++}(s,t,u)&=\frac{8G^2 s^4}{stu}\left[\frac{1}{\hat\ep}f_1(s,t,u)+f_2(s,t,u)\right]+4(G s)^2h(s,t,u)~,
\end{align}
where
\begin{align}
\label{240518.3}
f_1(s,t,u)&=s\ln(-s)+t\ln(-t)+u\ln(-u)~,\\
f_2(s,t,u)&=s\ln(-t)\ln(-u)+t\ln(-u)\ln(-s)+u\ln(-s)\ln(-t)~,\nn\\
h(s,t,u)&=\frac{1922 t^4+9143 t^3 u+14622 t^2 u^2+9143 t u^3+1922 u^4}{180 s^4}\nn\\
  & +\frac{(t+2u)(2t+u)(2t^4+2t^3u-t^2u^2+2tu^3+2u^4)}{s^6}\left(\ln^2\frac{t}{u}+\pi^2\right)\nn \nn \\
  &+\frac{(t-u)(341 t^4+1609 t^3u+2566t^2u^2+1609t u^3+341 u^4)}{30 s^5} \ln \frac{t}{u}~.\nn
\end{align}
For the other two helicity configurations 
the expressions for the one-loop scattering amplitudes are simpler,
\begin{align}
  \label{240425.7}
T^{(1)}_{+-;++}(s,t,u)&=\frac{G^2(s^2+t^2+u^2)}{90}~,\\
T^{(1)}_{--;++}(s,t,u)&=-\frac{G^2(s^2+t^2+u^2)}{30}~.\nn
\end{align}
All the scattering amplitudes explicitly displayed  are $t\leftrightarrow u$ symmetric because gravitons are Bosons and they appear with the same helicity in the final state. 
The general expression for $\Phi_{\beta\alpha}$ in Eq.~\eqref{240424.7}, that is independent of helicity, particularizes for graviton-graviton scattering as
\begin{align}
\label{240517.2}
\Phi(s,t,u)&=\frac{G}{\pi}\left(\frac{1}{\hat{\ep}}-\ln\Lambda^2\right)\left[s\ln(-s)+t\ln(-t)+u\ln(-u)\right]~.
\end{align}

Then, the infrared-finite one-loop scattering amplitude  $A^{(1)}_{++;++}(s,t,u)$, according to Eq.~\eqref{240423.3}, is given by 
\begin{align}
  \label{240424.10}
A_{++;++}^{(1)}(s,t,u)=T^{(1)}_{++;++}-T^{(0)}_{++;++}\Phi=\hat f_2(s,t,u)+4(Gs)^2h(s,t,u)\,, 
\end{align}
where,  for on-shell scattering $s+t+u=0$, it is straightforward to prove that\cite{Delgado:2022qnh}
\begin{align}
  \label{240424.11}
  \hat{f}_2(s,t,u)\equiv \ln\Lambda^2f_1(s,t,u)+f_2(s,t,u)&=s\ln\frac{-t}{\Lambda^2}\ln\frac{-u}{\Lambda^2}
  +t\ln\frac{-u}{\Lambda^2}\ln\frac{-s}{\Lambda^2}
  +u\ln\frac{-t}{\Lambda^2}\ln\frac{-s}{\Lambda^2}.
\end{align}
Importantly, now all the $\log$ functions appear with dimensionless arguments and the scattering amplitude can be directly used for meaningful computations, contrarily to $T^{(1)}_{++;++}(s,t,u)$ in Eq.~\eqref{240518.2}. The absence of infrared-divergences  in Eq.~\eqref{240425.7} reflects  the infrared-divergent structure given  in Eq.~\eqref{240423.3} \cite{Weinberg:1965nx}, which is zero at the one-loop level for a vanishing  tree-level amplitude, cf. Eq.~\eqref{240425.5}.

\subsection{Explicit formulas for the infrared-finite partial-wave amplitudes}

Let us discuss first the angular integration over the scattering angle $\int_{-1}^{+1}d\cos\theta\big(\cdots\big)$ of the real part of $A^{(1)}_{++;++}(s,t,u)$, Eq.~\eqref{240424.10}. Along the physical region, $s>0$, its imaginary part stems from $\ln(-s)=\ln(s)-i\pi$, with $s+i\vep$ and $\vep\to 0^+$.   In terms of the scattering angle $\theta$ and the variable $s$, one has the well-known formulas $t=-s \sin^2\frac{\theta}{2}$\,, and $u=-s \cos^2\frac{\theta}{2}$\,.  
The infrared-finite angular projection of $A_{++;++}^{(0)}$ was discussed in Ref.~\cite{Blas:2020och}, and we briefly review it below. It is clear that the angular integration for  $A^{(1)}_{+-;++}(s,t,u)$ and $A^{(1)}_{--;++}(s,t,u)$ does not pose any problem.  For the real part of $A_{++;++}^{(1)}(s,t,u)$  the only term that needs special discussion is the one divided by $stu$, since the rest of terms are polynomials in $t$ and $u$, sometimes multiplying $\ln t/u$ or $\ln^2 t/u$, which are integrable. 
Thus,  the limits in the integral that require further analysis  are $t\to 0$ (or $\theta\to 0$) and $u\to 0$ (or $\theta\to \pi$). Let us take $\theta\to 0$, for $\theta\to\pi$ one can proceed analogously with $t\leftrightarrow u$. Then, \jo{$u\to -s$ and} 
\begin{align}
  \label{240425.10}
s\ln\frac{-t}{\Lambda^2}\ln\frac{-u}{\Lambda^2}
  +t\ln\frac{-u}{\Lambda^2}\ln\frac{s}{\Lambda^2}
  +u\ln\frac{-t}{\Lambda^2}\ln\frac{s}{\Lambda^2}
  = {\cal O}(t \ln\frac{-t}{\Lambda^2})~,
\end{align}
so that when divided by $stu$ the angular integration is finite. Now, let us discuss the partial-wave projection for the imaginary part of $A^{(1)}_{++;++}(s,t,u)$. According to Eq.~\eqref{240424.10} we have that
\begin{align}
  \label{240425.11}
\Ima A^{(1)}_{++;++}(s,t,u)= -8\pi G^2s^3\left(\frac{1}{u}\ln\frac{-u}{\Lambda^2}+\frac{1}{t}\ln\frac{-t}{\Lambda^2}\right)~. 
\end{align}
In order to end with a meaningful infrared-finite partial-wave projection for $\Ima A_{++;++}^{(1)}(s,t,u)$, we have to consider Eq.~\eqref{240423.5} from the redefinition of the $S$ matrix. \jo{In this equation only the imaginary part of $\Phi$ enters in the exponent with its sign reversed. One can explicitly check from the  expression of $\Phi(s,t,u)$ given in Eq.~\eqref{240517.2}  that the real part of $\Phi$ vanishes for $\theta=0$ and $\pi$, which are the angles that enter in the normalization condition of two gravitons accompanying the identity operator. This explicitly proves that Eq.~\eqref{240423.5} is a consequence of the general formula Eq.~\eqref{240423.1a} that relates the infrared finite and diverging $S$ matrices. }

\jo{For two-graviton scattering} 
  $\Ima \Phi=-2Gs \ln\mu/\Lambda$ in the physical region, where $\mu\to 0^+$  is a graviton mass that acts as an infrared regulator, used in  Sec.~V of Ref.~\cite{Weinberg:1965nx} too.\footnote{At this point, this regularization is included to avoid having to treat with a partial-wave expansion in any dimension as it is the case with DR.} This generates an ${\cal O}(G^2)$ contribution to a PWA given by 
\begin{align}
  \label{240425.12}
i\frac{4G^2s^2}{\pi}\ln^2\left(\frac{\mu}{\Lambda} \right)\left(\delta_{\lambda'_1\lambda_1}\delta_{\lambda'_2\lambda_2}+(-1)^ J \delta_{\lambda'_1\lambda_2}\delta_{\lambda'_2\lambda_1}\right)~,  
\end{align}
It follows then that the complete formula for calculating a PWA for the one-loop scattering amplitude is
\begin{align}
\label{240519.2}
A^{J(1)}_{\lambda'_1\lambda'_2;\lambda_1\lambda_2}(s)&\!=\!
\frac{1}{8\pi^2}\!\!\int_{-1}^{+1}\!\! d\cos\theta  \,d_{\lam\lam'}^J(\theta)\,A^{(1)}_{\lambda'_1\lambda'_2;\lambda_1\lambda_2}(s,t,u)
+i\frac{4G^2s^2}{\pi}\ln^2\left(\frac{\mu}{\Lambda} \right)\left(\delta_{\lambda'_1\lambda_1}\delta_{\lambda'_2\lambda_2}+(-1)^ J \delta_{\lambda'_1\lambda_2}\delta_{\lambda'_2\lambda_1}\right)\,.
\end{align}
Here, $d^J_{\lam\lam'}(\theta)$ is the Wigner $d$-matrix, and $\lambda=\lambda_2-\lambda_1$, and $ \lambda'=\lambda'_2-\lambda'_1$~. 
  \jo{Let us exemplify Eq.~\eqref{240519.2}  for $J=0$ with $\lm_i=\lm'_i=+2$. It follows from this equation and Eq.~\eqref{240425.11}  that 
\begin{align}
\label{240425.14}
\Ima A^{0(1)}_{++;++}(s)&\!=\!\frac{8G^2s^2}{\pi}\ln^2\frac{\mu}{\Lambda}-\frac{2G^2s^3}{\pi}\int_{-1}^{+1}\!\!\! d\cos\theta
\frac{\ln(\frac{-t+\mu^2}{\Lambda^2})}{t-\mu^2}
\!=\!\frac{8G^2s^2}{\pi}\ln^2\frac{\mu}{\Lambda}+\frac{8G^2s^2}{\pi}\ln\frac{2p}{\mu}\ln\frac{2p\mu}{\Lambda^2}\!=\!\frac{8G^2s^2}{\pi}\ln^2\frac{2p}{\Lambda}~,
\end{align}
where, as an intermediate step we have regularized the integration in $\cos\theta$   by including a vanishing graviton mass, with the replacement $t\to t-\mu^2$. 
 Then, the partial-wave projection to ${\cal O}(G^2)$ becomes infrared finite. For $J>0$ one can proceed similarly to show that the PWA is IR finite by noticing that $P_J(1)=1$ \cite{Oller:2024qkw,Oller:2024nss}.}

For the LO case \cite{Blas:2020och,Blas:2020dyg}, the same process of accounting for the redefinition of the $S$ matrix at ${\cal O}(G)$ implies that
\begin{align}
  \label{240425.15}
A^{J(0)}_{\lambda'_1\lambda'_2;\lambda_1\lambda_2}(s)&=
\frac{1}{8\pi^2}
\int_{-1}^{+1}d\cos\theta \,d_{\lam\lam'}^J(\theta)\, A^{(0)}_{\lambda'_1\lambda'_2;\lambda_1\lambda_2}(s,t,u) 
+\frac{4G s}{\pi}\ln\!\left(\frac{\mu}{\Lambda}\right)\left(\delta_{\lambda'_1\lambda_1}\delta_{\lambda'_2\lambda_2}+(-1)^J \delta_{\lambda'_1\lambda_2}\delta_{\lambda'_2\lambda_1}\right).
\end{align}
 From Eqs.~\eqref{240519.2} and \eqref{240425.15} the resulting PWA is finite  and it also satisfies perturbative unitarity in PWAs at the one-loop level: 
\begin{align}
  \label{240425.13}
  \Ima A^{J(1)}_{\lambda'_1\lambda'_2;\lambda_1\lambda_2}(s)=\frac{\pi}{8}
\left(A^{J(0)}_{\lambda'_1\lambda'_2;\lambda_1\lambda_2}\right)^2\,,\,s\geq 0\,,  
\end{align} that was explicitly checked. \jo{For $J=0$ its fulfillment follows from Eqs.~\eqref{240425.14} and the formula $A^{0(0)}_{++;++}(s)=\frac{8Gs}{\pi}\ln\frac{2p}{\Lambda}$ that  one can directly calculate from Eq.~\eqref{240425.15}, cf.~Ref.~\cite{Blas:2020och}.} 

The analytical extrapolation in the complex $s$-plane of $A_{+-;++}^{(1)}(s,t,u)$ and $A^{(1)}_{--;++}(s,t,u)$, given in Eq.~\eqref{240425.7} is straightforward. For $A^{(1)}_{++;++}(s,t,u)$ 
the key point is to take into account the relation $\displaystyle{\ln\frac{-s}{\Lambda^2}=\ln\frac{s}{\Lambda^2}-i\sigma\pi}$\,, where $\sigma=\sgn\big( \Ima(s)\big)$ and $\arg(s)\in (-\pi,\pi)$.  
Then, it readily follows that for complex $s$ one can straightforwardly adapt  the arguments used above in regarding Eqs.~\eqref{240425.10} and \eqref{240425.11} (more details are given in Ref.~\cite{Oller:2024nss}). The neat result is that  for calculating the one-loop contribution to a PWA   from $A^{(1)}_{++;++}(s,t,u)$ in the complex $s$-plane one has the following formula:
\begin{align}
  \label{240430.4}
  A^{J(1)}_{++;++}(s)&=
    \frac{G^2s^2}{2\pi^2}    \int_{-1}^{+1}d\cos\theta\,P_J(\cos\theta)    \Bigg[\frac{2s}{tu}\left\{s\ln\frac{-t}{\Lambda^2}\ln\frac{-u}{\Lambda^2}+t\ln\frac{-u}{\Lambda^2}\ln\frac{s}{\Lambda^2}+u\ln\frac{-t}{\Lambda^2}\ln\frac{s}{\Lambda^2} \right\}+h(s,t,u) \Bigg] \nn\\
      &+i\sigma\frac{\pi}{8}\left(A^{J(0)}_{++;++}(s)\right)^2\,.
  \end{align}

Regarding the emerging scale $\Lambda$ we proceed as in Ref.~\cite{Blas:2020och} and write it as 
\begin{align}
  \label{241212.1}
  \Lambda^2=\frac{s}{a^2}
\end{align}
with $a>1$, since the approach of Ref.~\cite{Weinberg:1965nx} was developed for $\Lambda$  smaller (though not necessarily much smaller) than $s$.  Indeed, Weinberg in a couple of times refers in this way in Ref.~\cite{Weinberg:1965nx}. 
For the case of Coulomb scattering  $\ln a=\gamma_E$. This value can be derived rigorously in the non-relativistic case  \cite{Oller:2022tmo,Blas:2020dyg} by comparing with the known solution of Coulomb scattering in partial waves obtained in Quantum Mechanics. For QED the value is the same, as one can deduce through the comparison with Ref.~\cite{Bazhanov:1977fa}, which  employed the small-angle expansion of scattering amplitudes to all orders in the fine-structure constant.  As for graviton-graviton scattering, Ref.~\cite{Blas:2020dyg} argues that $\ln a\approx 1$ based on maximum smoothness in matching the graviball pole at $D=4$ to its pole position in larger dimensions, for which there are no infrared divergences \cite{Blas:2020dyg,Guerrieri:2022sod,Guerrieri:2021ivu}.

\jo{Let us emphasize that the relationship between the infrared-divergent and infrared-finite \( S \)-matrices in Eq.~\eqref{240423.1a} is established through the exponential factor \( \exp\left(\Phi_{\beta\alpha}\right) \), which has no poles. This allows us to use the new unitary and infrared-finite \( S \)-matrix \( S_{\beta\alpha}^{0} \) \cite{Weinberg:1965nx, Choi:2017bna} to search for its poles, thereby enabling the study of the related spectroscopy for graviton-graviton scattering. In close connection to this, it is important to note that the pole content of $S^0_{\beta\alpha}$ is independent of the value taken for the cutoff $\Lambda$, and hence of $a$, as it follows from Eq.~\eqref{241222.1}, since its right-hand side has no poles. Then, it is not either necessary for the cutoff \( \Lambda \) to satisfy \( \Lambda \ll \sqrt{s} \), nor for \( a \) to be excessively large (\( a \gg 1 \)), which clarifies the validity of our approach when \( \ln a \simeq 1 \).\footnote{The fact that $\ln a=\gamma_E$ for Coulomb scattering \cite{Oller:2022tmo,Bazhanov:1977fa} comes by considering the reproduction of the $S$ matrix with the standard phase choice.} 
 We also stress that this conclusion is independent of the  numerical value of \( G \) (its sign and strength), meaning that we can transition between different types of spectra, and the result stills hold.}

\section{Unitarized graviton-graviton scattering and the graviball}
\def\theequation{\arabic{section}.\arabic{equation}}
\setcounter{equation}{0}   
\label{sec.240517.2}

Given the PWA $A^{J}_{\lam'_1\lam'_2;\lam_1\lam_2}(s)$, its associated $S$ matrix is 
\begin{align}
\label{240519.3}
S^J_{\lam'_1\lam'_2;\lam_1\lam_2}(s)&=1+i\frac{\pi}{4}2^{|\lambda|/4}A^J_{\lam'_1\lam'_2;\lam_1\lam_2}(s)~.
\end{align}
For single-channel scattering, unitarity in PWAs implies that $|S^J_{\lam'_1\lam'_2;\lam_1\lam_2}(s)|^2=1$ which, at the level of the inverse of the PWA, reads \cite{Oller:1998hw}
\begin{align}
\label{240519.4}
\Ima\frac{1}{A^J_{\lam'_1\lam'_2;\lam_1\lam_2}(s)}=-\frac{\pi 2^{|\lam|/4}}{8}~,
\end{align}
valid for $s+i\vep$, with $s>0$. In this work we concentrate on studying the PWA $A^0_{++;++}(s)$, and more briefly $A^2_{++;++}(s)$ in the single-channel approximation.  We do not study here the impact of the PWAs $A^J_{--;++}(s)$, cf. Eq.~\eqref{240425.7}, which in principle couple to $A^J_{++;++}(s)$, because the former is very much suppressed ($A^{(1)}_{--;++}$ is reduced by a factor around 1500 compared to $A^{(1)}_{++;++}$).  
Results considering also PWAs built out of $A^{(1)}_{--;++}(s,t,u)$ will be shown in \cite{Oller:2024nss}.  Since we restrict in the following to helicities equal to $+2$ we drop them, and keep in mind that $\lambda=\lambda'=0$.

Taking advantage of Eq.~\eqref{240519.4}, one can fulfill unitarity by implementing a unitarization method, which should also respect proper analytical properties, like the existence of the unitarity or right-hand cut and of the left-hand cut. Two popular algebraic unitarization methods fulfilling these requirements are the IAM and the algebraic-$N/D$ methods.  
The IAM \cite{Lehmann:1972kv,Truong:1988zp,Dobado:1989qm,Delgado:2022uzu} can be straightforwardly obtained by performing the expansion of $1/A^J$ in powers of $G$ up to NLO, and then directly inverting the outcome. The PWA obtained is denoted by $A^J_{\tiam}(s)$ and is given by 
\begin{align}
\label{240519.5}
A^J_{\tiam}(s)&=\frac{A^{J(0)}(s)^2}{A^{J(0)}(s)-A^{J(1)}(s)}~.
\end{align}
$A^J_{\tiam}(s)$ fulfills unitarity, Eq.~\eqref{240519.4}, or equivalently that $\Ima A^J_{\tiam}(s)=|A^J_{\tiam}(s)|^2\pi/8$, as a consequence of $A^{J(1)}(s)$ satisfying  perturbative unitarity, Eq.~\eqref{240425.13}. 
The PWA resulting by the algebraic-$N/D$ method \cite{Oller:1998zr,Oller:1999me}, $A^J_{\tnd}(s)$, satisfies unitarity by implementing the parameterization
\begin{align}
\label{240519.6}
A^J_{\tnd}(s)=\frac{R^J(s)}{1+R^J(s)g(s)}~,
\end{align}
where $\displaystyle{g(s)=\frac{1}{8}\ln\frac{-s}{Q^2}}$.
We notice that $g(s)$ is an analytical function  
 with only a right-hand cut, \jo{and its imaginary part along the cut is $-\pi/8$, matching  $\Ima A^J(s)^{-1}$ in Eq.~\eqref{240519.4}. It follows then that   
 $\Ima R^J(s)=0$ for $s>0$, so that  it has only left-hand cut. This function is calculated perturbatively  in powers of $G$ by reproducing  the perturbative expansion of $A^J_{\tnd}(s)=A^{J(0)}+A^{J(1)}+{\cal O}(G^2)$ order by order. Thus,  $R^{J}=R^{J(0)}+R^{J(1)}+{\cal O}(G^2)$ with }
\begin{align}
\label{240519.8}
R^{J(0)}(s)&=A^{J(0)}(s)~,\\
\label{240519.9}
R^{J(1)}(s)&=A^{J(1)}(s)+A^{J(0)}(s)^2g(s)~.
\end{align}
\jo{At every order, the resulting function $R^J(s)$ is then substituted in Eq.~\eqref{240519.6}. Notice that $1/A^J(s)$  has the nice feature that it vanishes whenever  $A^{J}(s)$ (in the second Riemann sheet) has as resonance pole.}

Then, we distinguish between the application of the algebraic-$N/D$ method at LO, with $R^{J}(s)=R^{J(0)}(s)$, and at NLO, with $R^{J}=R^{J(0)}+R^{J(1)}$. The  former is denoted in short as ND0 and the latter as ND1, with the corresponding PWAs called $A^J_{\tndl}$ and $A^J_{\tndn}$, respectively.  
The unitarization scale $Q^2$ in $g(s)$, has an ultraviolet character. The effects of this scale are ${\cal O}(G^3)$ since, by construction, the algebraic-$N/D$ method reproduces the perturbative input to ${\cal O}(G^2)$.  Following the discussion in the introduction we take $Q^2=\pi/(G\ln a)$, and discuss results allowing a variation by $Q^2\to Q^2 2^\tau$, with $\tau=-1,0,1$. 
We have also explored the first-iterated $N/D$ method \cite{martin.200705.1,Oller:2020guq,Gulmez:2016scm}.
Nevertheless, we do not present its results here, since they are basically identical to those obtained by the algebraic method \cite{Oller:2024nss}. 
It is also worth indicating that the IAM can be seen as a particular case of the algebraic-$N/D$ method because
$\frac{1}{A^J_\tndn}=\frac{1}{A^{J(0)}}-\frac{A^{J(1)}}{(A^{J(0)})^2}+{\cal O}(G)= 
\frac{1}{A^J_{\tiam}}+{\cal O}\left(G\right).$

\begin{figure}
\begin{center}
\begin{tabular}{lll}
\hspace{-0.75cm}
\includegraphics[width=.33\textwidth, valign=c]{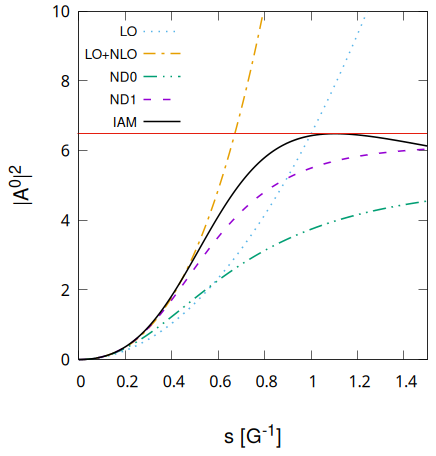} &  
\includegraphics[width=0.33\textwidth, valign=c]{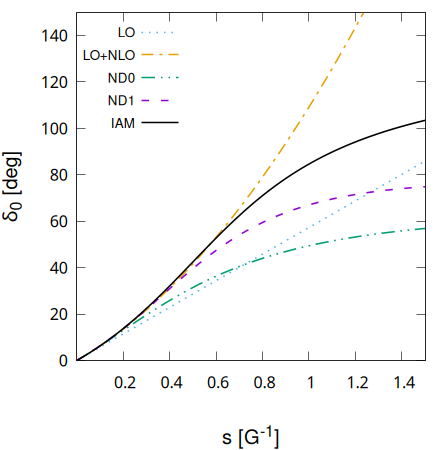} &
\begin{minipage}{12cm}
\hspace{-0.25cm}\includegraphics[width=.52\textwidth, valign=c]{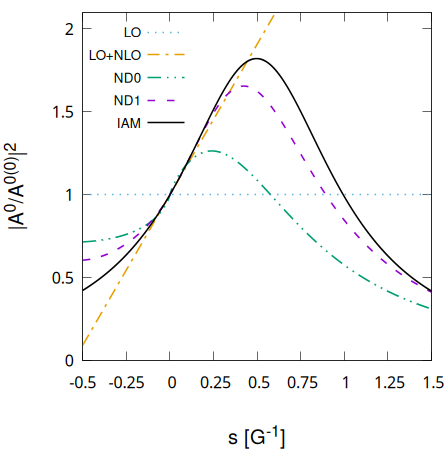} \end{minipage}
\end{tabular}
{\caption{\small In the left, center and right panels we plot  $\left|A^0(s)\right|^2$, the $S$-wave phase shifts $\delta_0(s)$, and $|A^0/A^{0(0)}|^2$. The perturbative calculations  up to LO and NLO correspond to the (blue) dotted and (orange) dashed-dotted lines, in this order. The results from the unitarization methods are: ND0  (green dashed-double-dotted lines), ND1 (magenta dashed lines), and the IAM (black solid lines).  For all calculations $\ln a=1$ is used, and  $Q^2=\pi /(G\ln a)$ is taken  for ND0 and ND1.}\label{fig.240520.1}}
\end{center}  \end{figure}

We first discuss the results for $\ln a=1$.  In the left and center panels of Fig.~\ref{fig.240520.1}   we show,  $|A^0(s)|^2$, and the phase shifts $\delta_0(s)$, respectively (for a PWA fulfilling unitarity $A^{0}(s)=\frac{8}{\pi}\sin\delta_0 e^{i\delta_0}$). They are obtained from the perturbative EFT calculations to LO (blue dotted) and NLO (orange dashed-dotted), and from $A^0_\tiam$ (black solid), $A^0_\tndl$ (green dashed-double-dotted), and $A^0_\tndn$ (magenta dashed line). \jo{We use $Q^2=\pi/(G\ln a)$ for ND0 and ND1. For the perturbative calculations} we find sizable NLO corrections already very soon for  $s\gtrsim 0.25~G^{-1}$. The perturbative result to NLO also violates the unitary bound for $|A^0|^2$, indicated by the horizontal red solid line in the left panel, for $s\gtrsim 0.6~G^{-1}$. It can neither reproduce  the typical bending in the unitarized results due to an underlying  resonance phenomenon.  The clear resonance shape in $|A^0/A^{0(0)}|^2$ for the unitarized results, shown in the right panel of Fig.~\ref{fig.240520.1}, is due to the presence of the graviball pole in the complex $s$ plane, which becomes more prominent for the NLO unitarized calculations. 

Seeking for  resonances implies the analytical extrapolation of a PWA to the second Riemann sheet (RS), which connects smoothly with the physical or first RS along the right-hand cut ($s>0$). The $T$ matrix in the second RS is denoted by $A^{J}_{\text{II}}(s)$, and resonances are poles of $A^{J}_{\text{II}}(s)$. The standard relation between $A^J(s)$ and $A^J_{\text{II}}(s)$ is 
$1/A^{J}_{\text{II}}=1/A^J+i\sigma \pi/4\,.$ 
The pole positions $s_P$ found for the different unitarization methods are shown in Table~\ref{tab.240502.1}, where $\ln a=1$ is used for all of them, and $Q^2=\pi/(G\ln a)$ is taken for ND0 and ND1. The LO study of Refs.~\cite{Blas:2020och,Blas:2020dyg} with ND0 finds a graviball pole at $s_{P;\tndl}=0.22-i\,0.64~G^{-1}$. Now, for the IAM the pole position is $s_{P;\tiam}=0.50-i\,0.55~G^{-1}$, so that  the real part  becomes larger and the imaginary part  smaller in absolute value compared with  $s_{P;\tndl}$. Because of this, the phase of $s_{P;\tiam}$ is 0.84, and that of $s_{P;\tndl}$ is  1.23, around a 30\% of relative difference. This size for a relative correction when passing from a LO to a NLO calculation indicates a suppression scale of around the value $Q^2=\pi G^{-1}$, such that $|s_p/Q^2|\approx 25\%$, in agreement with the estimates from the LO studies of Refs.~\cite{Blas:2020och,Blas:2020dyg}. Let us notice that $|s_P|$ only differs by a  8\% between both calculations, which is interesting because $|s_P|$ gives an upper bound for the scale expansion for the low-energy EFT of gravity.  For ND1 we have two poles. One pole corresponds to the graviball with $s_{P;\tndn}=0.42-i\,0.52~G^{-1}$,  with only a relative deviation of around a $10\%$ of difference with respect to $s_{P;\tiam}$, which matches with the expectation $|s_{P;\tiam}|^2/Q^2\simeq (0.75/\pi)^2\simeq 6\%$. 
In addition, ND1 has also another  pole  at $s_{Q;\tndn}=-0.87-i\,0.45~G^{-1}$. 
 This pole lies far away from the physical region $s>0$, so that it does not connect continuously with the physical sheet. 

  \begin{table}[h]
{\small  \begin{center}
    \begin{tabular}{lllll}
      Method & $s_P$ [$G^{-1}$] & $s_P$ [$\pi G^{-1}$] & Residue $\gamma^2~[G]$ & $\frac{\pi}{8}|\gamma^2|$~$[G^{-1}]$\\
      & $|s_P|$ [$G^{-1}$]  & $|s_P|$ [$\pi G^{-1}$] &$|\gamma^2|~[G]$ & \\
      \hline
      IAM    & $0.50-i\,0.55$  & $0.16-i\,0.17$ & $0.14+i\,1.40$ & 0.55 (0\%)\\
      &0.74 & 0.24 & 1.40 & \\
      \hline
ND0 & $0.22-i\,0.64$ & $0.07-i\,0.20$ & $1.10+i\,0.54$ & 0.48 (24\%)\\
& 0.68 & 0.22 & 1.23 & \\
ND1 & $0.42-i\,0.52$ & $0.13-i\,0.17$ & $0.38+i\,0.95$ & 0.40 (23\%)\\
& 0.67 & 0.21 & 1.02 &\\
\hline
\end{tabular}
\caption{{\small We show for each different method $s_P$ and the residue $\gamma^2$. 
In the last column we compare $|\Ima s_P|$ with $\pi|\gamma^2|/8$.  For all calculations $\ln a=1$ is used, and  $Q^2=\pi /(G\ln a)$ is taken for ND0 and ND1.
}\label{tab.240502.1}}
  \end{center}
}    \end{table}

The fact that $s_{P;\tiam}$ has a larger real part and a smaller imaginary part in modulus  than the LO determination \cite{Blas:2020och}, implies that  is closer to the real axis and  gives rise to a cleaner resonance signal. This is already seen in the more steep raise of the phase shifts $\delta_0$ for the IAM in the center panel of Fig.~\ref{fig.240520.1}, which also reach higher values passing through $\pi/2$, a point in which unitarity is saturated as shown in the left panel. A more clear way to see this resonance phenomenon is by depicting $\left|A^0/A^{0(0)}\right|^2$, shown in the right panel of Fig.~\ref{fig.240520.1}, which isolates better the resonance phenomenon by removing the zero of the PWA at $s=0$ \cite{Blas:2020och,Oller:2003ca}. As it can be seen, the results obtained by applying ND1 and IAM are very similar, while the perturbative  calculation to NLO 
matches well the unitarized results near $s=0$, but it clearly departs from the resonance shape due to the graviball for $|s|\gtrsim 0.25~G^{-1}$. The residue,  $\gamma^2$, of the PWA  at the pole position
  \begin{align}
    \label{240502.1}
    \gamma^2&=\lim_{s\to s_P}(s-s_P)A^{J=0}_{++;++}(s)
  \end{align}
 are shown in the fourth column of Table~\ref{tab.240502.1}.  The deviation between the moduli of the residues determined from ND0, ND1 and IAM are 
$(|\gamma^2_{\text{ND0}}|-|\gamma^2_{\text{ND1}}|)/\gamma^2_{\text{ND1}}=20\%$ and
$2(|\gamma^2_{\text{IAM}}|-|\gamma^2_{\text{ND1}}|)/(|\gamma^2_{\text{IAM}}|+|\gamma^2_{\text{ND1}}|)=31\%$. It is informative to compare with the narrow resonance case, for which 
$ -\Ima s_P= \frac{\pi}{8}|\gamma^2$.
The evaluation of $|\gamma^2|\pi/8$ for each pole is indicated in the fifth column of Table~\ref{tab.240502.1}, and the quantity between brackets indicates the relative deviation with respect to $|\Ima s_P|$. For the IAM the agreement is unexpectedly good, but for ND0 and ND1 the agreement is not perfect, as expected,  with a difference of around a 20\%.  We also mention that 
it was already shown in Ref.~\cite{Blas:2020dyg} that multi-graviton states, with more than two gravitons, are parametrically suppressed as possible coupled-channel intermediate states.  E.g. the phase space of three gravitons versus the one of two gravitons for $Gs=1$ was estimated to be suppressed by around $6\cdot 10^{-3}$.

\begin{figure}[h]
\begin{center}
\begin{tabular}{ll}
\begin{minipage}{8cm}
\vspace{-.2cm}
\includegraphics[width=.8\textwidth, valign=c]{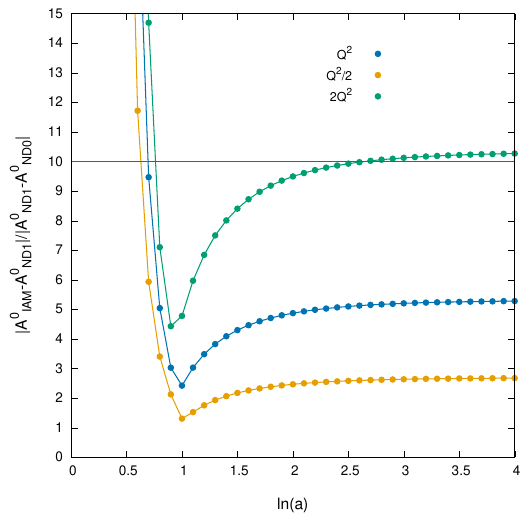}
\end{minipage} &
\begin{minipage}{6cm}\hspace{-0.6cm}\vspace{.32cm}\includegraphics[width=\textwidth,height=0.275\textheight,valign=c]{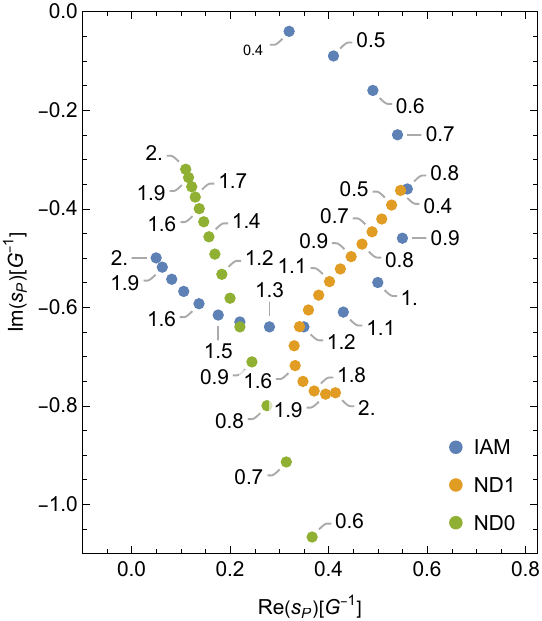} 
\end{minipage}
\end{tabular}
{\caption{\small $\ln a $ is varied in steps of 0.1. Left panel: Maximum of the ratio in Eq.~\eqref{240523.2}, for $s$ between 0 and $0.4 \,2^\tau Q^2$, with $\tau=0$, 1 and $-1$ for the blue, green and orange lines, respectively. Right panel:  Graviball poles for ND0 (green), IAM (blue), and ND1 (orange points). For every point the value of $\ln a$ used is indicated, and $Q^2=\pi/(G \ln a)$ is employed for ND0 and ND1.}\label{fig.240522.1} }
\end{center}  
\end{figure}

Let us now consider other values of $\ln a$. 
Since the difference in the results between the IAM and ND1 for $|s| \ll G^{-1}$ can be seen as a propagation of the error in $A^J_\tndn$ due to a variation $\delta R^{J(2)}$, we expect that  
\begin{align}
    \label{240520.2}
\left|A^J_\tiam-A^J_\tndn\right|=\left|\delta R^{J(2)}\left(\frac{A^J_\tndn}{R^J}\right)^2\right|~.
\end{align}
A similar reasoning for the variation between the results obtained by unitarizing the LO and NLO PWAs drives to
\begin{align}
    \label{240520.3}
\left|A^J_\tndl-A^J_\tndn\right|=\left|\delta R^{J(1)}\left(\frac{A^J_\tndn}{R^J}\right)^2\right|~.
\end{align}
Our final criterion is obtained by taking the quotient between Eqs.~\eqref{240520.2} and \eqref{240520.3}, so that the involved function $(A^J_\tndn/R^J)^2$ drops in the ratio. Taking into account that $\delta R^{J(2)}/\delta R^{J(1)}={\cal O}(s/Q^2)$, 
we  consider the maximum for the ratio 
\begin{align}
    \label{240523.2}
\left|\frac{(A^J_\tiam-A^J_\tndn)/(A^J_\tndn-A^J_\tndl)}{s/\widetilde{Q}^2}\right|
\end{align}
by varying $s$ from 0 up to $0.4\widetilde{Q}^2$, with $\widetilde{Q}^2=2^\tau Q^2$
and $\tau=-1,0,1$, {\it i.e.} for   $Q^2/2$, $Q^2$, and $2 Q^2$, respectively.  
We show in the left panel of Fig.~\ref{fig.240522.1} the result of this exercise as a function of $\ln a$ in steps of $0.1$, with the colors blue, orange and green corresponding to $\tau=0,~1,$ and $-1$, respectively.  
We see that this ratio exceedingly grows for $\ln a\lesssim 0.5$, which is a clear hint that this parametric region of small values of $\ln a$ should be disregarded on the basis of an acceptable behavior of the unitarized EFT. For the rest of values shown in the figure, we observe that there is a minimum for $\ln a\approx 1$,  and then the ratio  grows and it finally keeps rather constant. 
We can also interpret the value at the minimum as an indication that it does not make much sense to decrease(increase) further the unitarization scale beyond $\frac{Q^2}{2}(2Q^2)$, because otherwise this minimum value would be too small(large) compared with expectations. 

\begin{figure}
\begin{center}
\begin{tabular}{ll}
\includegraphics[width=.45\textwidth, valign=c]{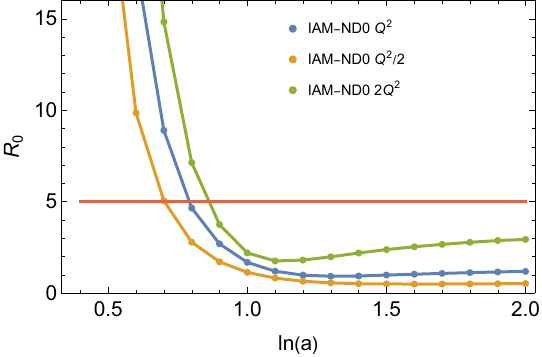} & \includegraphics[width=.45\textwidth, valign=c]{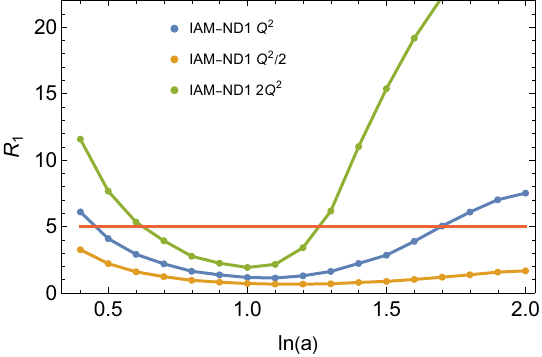} 
\end{tabular}
{\caption{\small The ratios $R_0$ (left) and $R_1$ (right panel) are shown, cf. Eqs.~\eqref{240524.1} and \eqref{240524.2}, respectively, as a function of $\ln a$ in steps of 0.1.  The results from three different unitarization scales  used in ND0 and ND1 are plotted: $Q^2$ (blue), $Q^2/2$ (orange) and $2Q^2$ (green).}\label{fig.240522.2a} }
\end{center}  
\end{figure}

We show in the right panel of Fig.~\ref{fig.240522.1} the graviball poles, with every point labelled with its corresponding value of $\ln a$. 
From the figure it is clear that the poles from the different methods lie closer when $\ln a$ is roughly in between 0.8 and 1.5. There $s_{P;\tiam}$ moves from right to left in between $0.55-i\,0.35$ to $0.15-i\,0.6~G^{-1}$. While this visually pole-accumulating region is appealing, one can quantify what is meant by closer and further by measuring  distances  with respect to $Q^2$, which  scales like $1/\ln a$.  For that aim we take the ratios 
\begin{align}
    \label{240524.1}
R_{0}&=\left|\frac{s_{P;\tiam}-s_{P;\tndl}}{s_{P;\tiam}}\right|
/\left|\frac{s_{P;\tiam}}{2^\tau Q^2}\right|~,\\
\label{240524.2}
R_{1}&=\left|\frac{s_{P;\tiam}-s_{P;\tndn}}{s_{P;\tiam}-s_{P;\tndl}}\right|/\left|\frac{s_{P;\tiam}}{2^\tau Q^2}\right|~.
\end{align}
These ratios are plotted  in the left and right panels of Fig.~\ref{fig.240522.2a}. Let us notice that the positions for all poles used in this figure are smaller in modulus than $2^\tau Q^2$, so that an order of magnitude estimate based on powers of $s/(2^\tau Q^2)$ is in order. 
  We see that small values of $\ln a\lesssim 0.5$ are also disfavored from the study of these ratios, in agreement with our previous observations regarding Fig.~\ref{fig.240522.1}.  
  In addition, the ratio $R_1$ hints to disfavor larger values of $\ln a\gtrsim 1.7$ because, out of the three scales, the two larger ones provide  large $R_1 >5$.  This interval of values for $\ln a$ 
  is similar to the preferred one  in between 0.8 and 1.3  found in Ref.~\cite{Blas:2020dyg} from an independent dimensional analysis. 

\jo{Let us elaborate on the meaning of the optimized interval of values for $\ln a$ around 1 within our method. 
  From the point of view of EFT, we already remarked above that the parameter $\ln a$  controls the size of the ultraviolet scale $Q^2\simeq \pi G/\ln a$, that we determine by studying the unitarity corrections \cite{Blas:2020dyg,Blas:2020och} from the combination $A^{J(0)}_{++;++} g(s)\to s\frac{G \ln a}{\pi}\equiv s/Q^2$.  The corrections from the imaginary part of $g(s)$ are larger by a factor $\pi$ at NLO.  Within the unitarized theory  the larger corrections from $\text{Im}g(s)$ are immediately accounted for, as well as a calculation of those unitarity corrections from the real part of the unitarity loops  by invoking analyticity along the right-hand cut. 
  From the EFT perspective, it does not make sense for \( \ln a \) to be too large, as this would make $Q^2$ too small, compromising the convergence properties of the perturbative expansion needed for the calculation of \( R^J \), cf.  Eqs.~\eqref{240519.8} and \eqref{240519.9}. On the other hand, if \( \ln a \) were too small, \( Q^2 \) would be an overly optimistic (too large) estimate for the actual ultraviolet scale.  Thus, from the EFT viewpoint, \( \ln a \) should be \( {\cal O}(1) \); otherwise, the situation could be considered fine tuned. Our analysis, which considers corrections from LO to NLO and compares different unitarization methods, provides a more accurate determination of the optimal interval for \( \ln a \), which is \( 0.5 \lesssim \ln a \lesssim 1.7 \). In general, as the order of calculation increases, the dependence of the poles on $\ln a$ for $\ln a = {\cal O}(1)$ is expected to decrease, while the range of values for which this holds should expand and become broader. This trend is clearly illustrated in Fig.~\ref{fig.240522.1}: the pole positions for ND1 (orange points) are  clustered within a remarkably narrow region, whereas the pole positions for ND0 at LO remain unbounded as $\ln a \to 0^+$.}

Let us now briefly discuss the $J=2$ PWA $A^2_{++;++}(s)$. The calculated infrared-finite perturbative PWA up to NLO satisfies perturbative unitarity. 
For limitations of space we do not show here the results for $A^2(s)$ from the unitarization methods with $\ln a= 1$. The resulting $J=2$  PWA is clearly  more perturbative than the $J=0$ already discussed, with higher order corrections becoming relevant for $s\gtrsim 0.6~G^{-1}$. We also notice that $\delta_2(s)<0$, which indicates the repulsive character of the $D$-wave graviton-graviton interaction in this energy region. This is also in agreement with the fact that  now for $D$-wave scattering we do not find any resonance pole in the second RS that could impact the physical region of interest. The perturbative LO PWA is given by  $A^{2(0)}=-4s(3-2\ln a)/\pi$, and for $\ln a=3/2$  changes sign, so that the $D$-wave interaction passes from repulsive to attractive. 
For larger $\ln a>2.5$ the attractive interaction has enough strength to give rise to resonance poles with $|s_P|<1~G^{-1}$, that become lighter as $\ln a$ further increases. However, this parametric region for $\ln a$ has been already disfavored and,  in fact, the $D$-wave pole positions vary then very appreciably from one unitarization  method to another.

\section{Conclusions and outlook}
\def\theequation{\arabic{section}.\arabic{equation}}
\setcounter{equation}{0}
\label{sec.240517.4}

Summarizing, we have discussed how to calculate infrared-finite partial-wave amplitudes (PWAs) for graviton-graviton scattering, and apply it to  one loop in perturbative gravity \cite{Dunbar:1994bn}. For removing the infrared divergences we have first followed the results of Ref.~\cite{Weinberg:1965nx}. However, when performing the angular integrals needed to calculate a PWA new divergences arise. To remove them, one has to account for a redefinition  by a  phase factor \cite{Blas:2020och} of the identity operator, which is part of the relation between the $S$- and $T$-matrices. Remarkably, the resulting  PWAs satisfy perturbative unitarity. \jo{The removal of the infrared divergences gives rise to the emergence of a dimensionless parameter $\ln a$, such that the cutoff squared in momentum  is expressed as $\Lambda^2=s/a^2$, which differentiates between soft(hard) gravitons having linear momenta smaller(larger) than $\Lambda$ in magnitude. We have also shown that the spectrum of the theory (poles in the PWAs) does not depend on the specific value of $\ln a$ in a calculation to all orders in powers of $G$.}

In order to end with nonperturbative PWAs that satisfy full unitarity and can be used to search for resonances, the  perturbative PWAs are unitarized by applying the Inverse Amplitude Method (IAM) and the algebraic-$N/D$ method. These methods have been extensively employed \jo{with success} in hadron physics \cite{Oller:2020guq,Oller:2019opk}. We have then studied in detail the $J=0$ graviton-graviton PWA. As a result of unitarizing this PWA  the existence of the graviball pole, predicted by the unitarization of the Born amplitude in Ref.~\cite{Blas:2020och}, is confirmed, with a pole  position that is similar for the different unitarization methods. With the IAM no unitarization scale is introduced,   while one is needed for the algebraic $N/D$ method and, in particular, for the leading-order study of \cite{Blas:2020och}.  
The requirement of ending with a well-behaved unitarized effective-field theory (EFT) implies interesting constraints on $\ln a$. In this respect, we have compared the variation in the results between the different unitarization methods, and between the unitarized results to LO and NLO order. 
We conclude that small values of $\ln a\lesssim 0.5$ should be discarded, because  variations observed are too large  compared with expectations. Similarly, hints for discarding  larger values of $\ln a\gtrsim 1.7$ are clear too. \jo{Within the  range $0.5\lesssim \ln a\lesssim 1.7$ the graviball pole position $s_P$ from the algebraic-$N/D$ method to NLO  varies approximately from $0.5-i\,0.4~$G$^{-1}$ to $0.3-i\,0.7$~G$^{-1}$\,. A variation much smaller than the one from the LO study.} 
We have also studied the $D$-wave graviton-graviton scattering. For  $\ln a\approx 1$ it is clearly weaker than its $S$-wave counterpart,   it is repulsive and  not resonant. 

We expect to further pursue these techniques in the future to two-loop graviton-graviton scattering calculated in perturbative gravity \cite{Abreu:2020lyk}. It will be also exciting their extensive application  to systems made up not only by gravitons but also by  matter fields. As the number of light fields coupled increases, it is expected that the graviball becomes lighter and narrower \cite{Blas:2020dyg}. So, the exciting prospect that the graviball could be relevant for physics at much lower scales, e.g. those in inflation or in theories with extra dimensions \cite{Arkani-Hamed:1998jmv,Antoniadis:1998ig}, should be clarified. It would be also interesting to study the connection between the graviball and the  portrait of a black hole as a leaking quantum Bose condensate of soft gravitons \cite{Dvali:2010bf,Dvali:2011aa}.

\section*{Acknowledgements}
JAO would like to acknowledge partial financial support to the Grant PID2022-136510NB-C32 funded
by \\MCIN/AEI/10.13039/501100011033/ and FEDER, UE, and to the EU Horizon 2020 research and innovation program, STRONG-2020 project, under grant agreement no. 824093. MP acknowledges financial support from the ANII-FCE-175902 project.

\bibliographystyle{unsrt}
\bibliography{references}

\begin{thebibliography}{10}

\bibitem{Dunbar:1994bn}
David~C. Dunbar and Paul~S. Norridge.
\newblock {Calculation of graviton scattering amplitudes using string based
  methods}.
\newblock {\em Nucl. Phys. B}, 433:181--208, 1995.

\bibitem{vanNieuwenhuizen:1976vb}
P.~van Nieuwenhuizen and C.~C. Wu.
\newblock {On Integral Relations for Invariants Constructed from Three Riemann
  Tensors and their Applications in Quantum Gravity}.
\newblock {\em J. Math. Phys.}, 18:182, 1977.

\bibitem{AccettulliHuber:2019jqo}
Manuel Accettulli~Huber, Andreas Brandhuber, Stefano De~Angelis, and Gabriele
  Travaglini.
\newblock {Note on the absence of $R^2$ corrections to Newton\textquoteright{}s
  potential}.
\newblock {\em Phys. Rev. D}, 101(4):046011, 2020.

\bibitem{Dunbar:1995ed}
David~C. Dunbar and Paul~S. Norridge.
\newblock {Infinities within graviton scattering amplitudes}.
\newblock {\em Class. Quant. Grav.}, 14:351--365, 1997.

\bibitem{Donoghue:1999qh}
John~F. Donoghue and Tibor Torma.
\newblock {Infrared behavior of graviton-graviton scattering}.
\newblock {\em Phys. Rev. D}, 60:024003, 1999.

\bibitem{Donoghue:2017pgk}
John~F. Donoghue, Mikhail~M. Ivanov, and Andrey Shkerin.
\newblock {EPFL Lectures on General Relativity as a Quantum Field Theory}.
\newblock arXiv:1702.00319 [hep-th] 2017.

\bibitem{Weinberg:1978kz}
Steven Weinberg.
\newblock {Phenomenological Lagrangians}.
\newblock {\em Physica A}, 96(1-2):327--340, 1979.

\bibitem{Donoghue:1993eb}
John~F. Donoghue.
\newblock {Leading quantum correction to the Newtonian potential}.
\newblock {\em Phys. Rev. Lett.}, 72:2996--2999, 1994.

\bibitem{Donoghue:1994dn}
John~F. Donoghue.
\newblock {General relativity as an effective field theory: The leading quantum
  corrections}.
\newblock {\em Phys. Rev. D}, 50:3874--3888, 1994.

\bibitem{Bjerrum-Bohr:2002gqz}
N.~E.~J Bjerrum-Bohr, John~F. Donoghue, and Barry~R. Holstein.
\newblock {Quantum gravitational corrections to the nonrelativistic scattering
  potential of two masses}.
\newblock {\em Phys. Rev. D}, 67:084033, 2003.
\newblock [Erratum: Phys.Rev.D 71, 069903 (2005)].

\bibitem{Burgess:2003jk}
C.~P. Burgess.
\newblock {Quantum gravity in everyday life: General relativity as an effective
  field theory}.
\newblock {\em Living Rev. Rel.}, 7:5--56, 2004.

\bibitem{Ware:2013zja}
John Ware, Ryo Saotome, and Ratindranath Akhoury.
\newblock {Construction of an asymptotic S matrix for perturbative quantum
  gravity}.
\newblock {\em JHEP}, 10:159, 2013.

\bibitem{Choi:2017bna}
Sangmin Choi, Uri Kol, and Ratindranath Akhoury.
\newblock {Asymptotic Dynamics in Perturbative Quantum Gravity and BMS
  Supertranslations}.
\newblock {\em JHEP}, 01:142, 2018.

\bibitem{Choi:2017ylo}
Sangmin Choi and Ratindranath Akhoury.
\newblock {BMS Supertranslation Symmetry Implies Faddeev-Kulish Amplitudes}.
\newblock {\em JHEP}, 02:171, 2018.

\bibitem{Prabhu:2022zcr}
Kartik Prabhu, Gautam Satishchandran, and Robert~M. Wald.
\newblock {Infrared finite scattering theory in quantum field theory and
  quantum gravity}.
\newblock {\em Phys. Rev. D}, 106(6):066005, 2022.

\bibitem{Prabhu:2024zwl}
Kartik Prabhu and Gautam Satishchandran.
\newblock {Infrared finite scattering theory: scattering states and
  representations of the BMS group}.
\newblock {\em JHEP}, 08:055, 2024.

\bibitem{Prabhu:2024lmg}
Kartik Prabhu and Gautam Satishchandran.
\newblock {Infrared finite scattering theory: Amplitudes and soft theorems}.
\newblock {\em Phys. Rev. D}, 110(8):085022, 2024.

\bibitem{Campiglia:2016hvg}
Miguel Campiglia and Alok Laddha.
\newblock {Subleading soft photons and large gauge transformations}.
\newblock {\em JHEP}, 11:012, 2016.

\bibitem{Campiglia:2014yka}
Miguel Campiglia and Alok Laddha.
\newblock {Asymptotic symmetries and subleading soft graviton theorem}.
\newblock {\em Phys. Rev. D}, 90(12):124028, 2014.

\bibitem{Ashtekar:2018lor}
Abhay Ashtekar, Miguel Campiglia, and Alok Laddha.
\newblock {Null infinity, the BMS group and infrared issues}.
\newblock {\em Gen. Rel. Grav.}, 50(11):140--163, 2018.

\bibitem{Hirai:2020kzx}
Hayato Hirai and Sotaro Sugishita.
\newblock {IR finite S-matrix by gauge invariant dressed states}.
\newblock {\em JHEP}, 02:025, 2021.

\bibitem{Himwich:2020rro}
Elizabeth Himwich, Sruthi~A. Narayanan, Monica Pate, Nisarga Paul, and Andrew
  Strominger.
\newblock {The Soft $\mathcal{S}$-Matrix in Gravity}.
\newblock {\em JHEP}, 09:129, 2020.

\bibitem{Hannesdottir:2019umk}
Holmfridur Hannesdottir and Matthew~D. Schwartz.
\newblock {Finite $S$ matrix}.
\newblock {\em Phys. Rev. D}, 107(2):L021701, 2023.

\bibitem{Elkhidir:2024izo}
Asaad Elkhidir, Donal O'Connell, and Radu Roiban.
\newblock {Supertranslations from Scattering Amplitudes}.
\newblock arXiv:2408.15961~[hep-th], 2024.

\bibitem{Kulish:1970ut}
P.~P. Kulish and L.~D. Faddeev.
\newblock {Asymptotic conditions and infrared divergences in quantum
  electrodynamics}.
\newblock {\em Theor. Math. Phys.}, 4:745, 1970.

\bibitem{Oller:2024qkw}
J.~A. Oller and Marcela Pel\'aez.
\newblock {Unitarization of electron scattering with an external potential at
  NLO in QED}.
\newblock {\em JHEP}, 11:113, 2024.

\bibitem{Weinberg:1965nx}
Steven Weinberg.
\newblock {Infrared photons and gravitons}.
\newblock {\em Phys. Rev.}, 140:B516--B524, 1965.

\bibitem{Blas:2020dyg}
Diego Blas, Jorge Martin~Camalich, and J.~A. Oller.
\newblock {Unitarization of infinite-range forces: graviton-graviton
  scattering}.
\newblock {\em JHEP}, 08:266, 2022.

\bibitem{Blas:2020och}
D.~Blas, J.~Martin~Camalich, and J.~A. Oller.
\newblock {Scalar resonance in graviton-graviton scattering at high-energies:
  The graviball}.
\newblock {\em Phys. Lett. B}, 827:136991, 2022.

\bibitem{Oller:2022tmo}
J.~A. Oller.
\newblock {Unitarizing non-relativistic Coulomb scattering}.
\newblock {\em Phys. Lett. B}, 835:137568, 2022.

\bibitem{Oller:2020guq}
J.~A. Oller.
\newblock {Unitarization Technics in Hadron Physics with Historical Remarks}.
\newblock {\em Symmetry}, 12(7):1114, 2020.

\bibitem{Oller:2019opk}
J.~A. Oller.
\newblock {Coupled-channel approach in hadron\textendash{}hadron scattering}.
\newblock {\em Prog. Part. Nucl. Phys.}, 110:103728, 2020.

\bibitem{Oller:1998zr}
J.~A. Oller and E.~Oset.
\newblock {N/D description of two meson amplitudes and chiral symmetry}.
\newblock {\em Phys. Rev. D}, 60:074023, 1999.

\bibitem{Oller:1999me}
J.~A. Oller.
\newblock {The Case of a $W W$ dynamical scalar resonance within a chiral
  effective description of the strongly interacting Higgs sector}.
\newblock {\em Phys. Lett. B}, 477:187--194, 2000.

\bibitem{martin.200705.1}
A.~D. Martin and T.~D. Spearman.
\newblock {\em Elementary Particle Theory}.
\newblock North-Holland Publishing Company, Amsterdam, 1970.

\bibitem{Lehmann:1972kv}
H.~Lehmann.
\newblock {Chiral invariance and effective range expansion for pion pion
  scattering}.
\newblock {\em Phys. Lett. B}, 41:529--532, 1972.

\bibitem{Truong:1988zp}
Tran~N. Truong.
\newblock {Chiral Perturbation Theory and Final State Theorem}.
\newblock {\em Phys. Rev. Lett.}, 61:2526, 1988.

\bibitem{Dobado:1989qm}
A.~Dobado, Maria~J. Herrero, and Tran~N. Truong.
\newblock {Unitarized Chiral Perturbation Theory for Elastic Pion-Pion
  Scattering}.
\newblock {\em Phys. Lett. B}, 235:134--140, 1990.

\bibitem{Delgado:2022uzu}
Rafael~L. Delgado, Antonio Dobado, and Dom\`enec Espriu.
\newblock {Unitarized one-loop graviton-graviton scattering}.
\newblock {\em EPJ Web Conf.}, 274:08010, 2022.

\bibitem{Oller:2019rej}
J.~A. Oller.
\newblock {\em {A Brief Introduction to Dispersion Relations}}.
\newblock SpringerBriefs in Physics. Springer, 2019.

\bibitem{Oller:2000ma}
J.~A. Oller, E.~Oset, and A.~Ramos.
\newblock {Chiral unitary approach to meson meson and meson - baryon
  interactions and nuclear applications}.
\newblock {\em Prog. Part. Nucl. Phys.}, 45:157--242, 2000.

\bibitem{Mai:2020ltx}
Maxim Mai.
\newblock {Review of the ${\Lambda }$(1405) A curious case of a strangeness
  resonance}.
\newblock {\em Eur. Phys. J. ST}, 230(6):1593--1607, 2021.

\bibitem{ParticleDataGroup:2022pth}
R.~L. Workman et~al.
\newblock {Review of Particle Physics}.
\newblock {\em PTEP}, 2022:083C01, 2022.

\bibitem{Weinberg:1966fm}
Steven Weinberg.
\newblock {Dynamical approach to current algebra}.
\newblock {\em Phys. Rev. Lett.}, 18:188--191, 1967.

\bibitem{Gasser:1983yg}
J.~Gasser and H.~Leutwyler.
\newblock {Chiral Perturbation Theory to One Loop}.
\newblock {\em Annals Phys.}, 158:142, 1984.

\bibitem{Oller:1997ti}
J.~A. Oller and E.~Oset.
\newblock {Chiral symmetry amplitudes in the S wave isoscalar and isovector
  channels and the $\sigma$, f$_0$(980), a$_0$(980) scalar mesons}.
\newblock {\em Nucl. Phys. A}, 620:438--456, 1997.
\newblock [Erratum: Nucl.Phys.A 652, 407--409 (1999)].

\bibitem{Caprini:2005zr}
Irinel Caprini, Gilberto Colangelo, and Heinrich Leutwyler.
\newblock {Mass and width of the lowest resonance in QCD}.
\newblock {\em Phys. Rev. Lett.}, 96:132001, 2006.

\bibitem{Garcia-Martin:2011nna}
R.~Garcia-Martin, R.~Kaminski, J.~R. Pelaez, and J.~Ruiz~de Elvira.
\newblock {Precise determination of the f0(600) and f0(980) pole parameters
  from a dispersive data analysis}.
\newblock {\em Phys. Rev. Lett.}, 107:072001, 2011.

\bibitem{Oller:2024nss}
J.~A. Oller and M.~Pel\'aez.
\newblock {{\it In preparation}}.
\newblock 2025.

\bibitem{Delgado:2022qnh}
Rafael~L. Delgado, Antonio Dobado, and Dom\`enec Espriu.
\newblock {Seeking for resonances in unitarized one-loop graviton-graviton
  scattering}.
\newblock {\em Phys. Rev. D}, 107(4):044073, 2023.

\bibitem{Bazhanov:1977fa}
V.~V. Bazhanov, G.~P. Pronko, L.~D. Solovev, and Yu.~Ya. Yushin.
\newblock {Small Angle Scattering in Quantum Electrodynamics}.
\newblock {\em Teor. Mat. Fiz.}, 33:218--230, 1977.

\bibitem{Guerrieri:2022sod}
Andrea Guerrieri, Harish Murali, Joao Penedones, and Pedro Vieira.
\newblock {Where is M-theory in the space of scattering amplitudes?}
\newblock {\em JHEP}, 06:064, 2023.

\bibitem{Guerrieri:2021ivu}
Andrea Guerrieri, Joao Penedones, and Pedro Vieira.
\newblock {Where Is String Theory in the Space of Scattering Amplitudes?}
\newblock {\em Phys. Rev. Lett.}, 127(8):081601, 2021.

\bibitem{Oller:1998hw}
J.~A. Oller, E.~Oset, and J.~R. Pelaez.
\newblock {Meson meson interaction in a nonperturbative chiral approach}.
\newblock {\em Phys. Rev. D}, 59:074001, 1999.
\newblock [Erratum: {\it ibid} 60, 099906 (1999); {\it ibid} 75, 099903
  (2007)].

\bibitem{Gulmez:2016scm}
D.~G\"ulmez, U.~G. Mei\ss{}ner, and J.~A. Oller.
\newblock {A chiral covariant approach to $\rho\rho$ scattering}.
\newblock {\em Eur. Phys. J. C}, 77(7):460, 2017.

\bibitem{Oller:2003ca}
J.~A. Oller.
\newblock {On the resonant and nonresonant contributions to $B \to \rho\pi$}.
\newblock {\em eConf}, C0304052:WG412, 2003.

\bibitem{Abreu:2020lyk}
S.~Abreu, F.~Febres~Cordero, H.~Ita, M.~Jaquier, B.~Page, M.S. Ruf, and
  V.~Sotnikov.
\newblock {The Two-Loop Four-Graviton Scattering Amplitudes}.
\newblock {\em Phys. Rev. Lett.}, 124(21):211601, 2020.

\bibitem{Arkani-Hamed:1998jmv}
Nima Arkani-Hamed, Savas Dimopoulos, and G.~R. Dvali.
\newblock {The Hierarchy problem and new dimensions at a millimeter}.
\newblock {\em Phys. Lett. B}, 429:263--272, 1998.

\bibitem{Antoniadis:1998ig}
Ignatios Antoniadis, Nima Arkani-Hamed, Savas Dimopoulos, and G.~R. Dvali.
\newblock {New dimensions at a millimeter to a Fermi and superstrings at a
  TeV}.
\newblock {\em Phys. Lett. B}, 436:257--263, 1998.

\bibitem{Dvali:2010bf}
Gia Dvali and Cesar Gomez.
\newblock {Self-Completeness of Einstein Gravity}.
\newblock arXiv:1005.3497 [hep-th].

\bibitem{Dvali:2011aa}
Gia Dvali and Cesar Gomez.
\newblock {Black Hole's Quantum N-Portrait}.
\newblock {\em Fortsch. Phys.}, 61:742--767, 2013.

\end{thebibliography}

\end{document}